\documentclass[]{aa} 

\usepackage{txfonts}
\usepackage{graphicx}
\usepackage[authoryear]{natbib}

\usepackage{longtable} 

\bibpunct{(}{)}{;}{a}{}{,} 

\newcommand{\Teff}{\ensuremath{T_{\mathrm{eff}}}}
\newcommand{\logg}{\ensuremath{\log g}}

\begin{document}

\title{Measuring helium abundance difference in giants of NGC 2808   
\thanks{Based on observations collected at ESO VLT, Chile, under programme 
384.D-0283.}
} 
  

\author{L. Pasquini\inst{1} \and P. Mauas\inst{2}  \and H.U. K\"aufl\inst{1} 
\and C. Cacciari\inst{3}   }

\offprints{L. Pasquini, \email{lpasquin@eso.org}}

\institute{ESO -- European Southern Observatory, Karl-Schwarzschild-Strasse 2, 85748 Garching bei M\"unchen, Germany
 \and IAFE, CONICET-UBA and FCEN, UBA, Ciudad Universitaria, Buenos Aires, Argentina   
  \and Istituto Nazionale di Astrofisica, Osservatorio Astronomico di Bologna, 
  Via Ranzani 1,  40127 Bologna, Italy
}

\date{Received  / Accepted }

\abstract
{Multiple populations have been  detected in several globular clusters (GC) that do  
not display a spread in metallicity. 
Unusual features of their observed colour-magnitude diagrams (CMD) can be interpreted in terms of
 differences in the Helium content of the stars belonging to the sub-populations.  
} 
{ Even if evidence gathered so far is compelling, differences in He abundance have never 
been directly observed. We attempt to  measure these differences in two giant stars of NGC 2808 
with very similar astrophysical parameters but different Na and O abundances, hence that presumably 
belong to different sub-populations, by directly comparing their He I 10830 {\AA} lines.  
}  
{The He 10830 {\AA} line forms in the upper chromosphere. Our detailed models derive the 
chromospheric structure using the Ca II and H$\alpha$ chromospheric lines, and simulate the 
corresponding He I 10830 line profiles as a function of the helium abundance. We show that, 
at a given value of He abundance, the He I 10830 equivalent width cannot significantly 
change without a corresponding much larger change in the Ca II  chromospheric lines.
We have used the VLT-CRIRES to obtain high-resolution spectra in the 10830 {\AA} region, and the 
VLT-UVES to obtain spectra of the Ca II and  H$\alpha$ lines of our target stars.
}    
{The two target stars have very similar Ca II and H$\alpha$ chromospheric lines, but different 
appearances in the He region.  
One line, blueshifted by 17 km~s$^{-1}$ with respect to the He 10830 rest wavelength, 
is detected in the spectrum of the 
Na-rich star, whereas the  Na-poor star spectrum is consistent with a non-detection. 
From a detailed chromospheric modeling, we show that the difference in 
the spectra is consistent and most closely explained by an  He abundance difference between the two stars
of  $\Delta Y \ge$ 0.17. Our optical observations bracket the infrared ones over a range of 
about 50 days and we do not observe any substantial variability
in the Ca II and H$\alpha$ lines. 
} 
{  We provide direct evidence of a significant He line strength difference
in giant stars of NGC 2808 belonging to different sub-populations, which had been 
previously detected by other photometric and spectroscopic means. 
}

   
\titlerunning{Measurement of He variations in GC RGB stars}
\authorrunning{L. Pasquini et al.}
\maketitle

\section{Introduction}
\label{s:intro}
Light element (CNO cycle elements, Mg, Na) variations and anti-correlations 
in globular cluster (GC) red giant branch (RGB) stars have been known to exist for 
nearly 20 years (see e.g. Carretta et al. 2010 for a review) and for a long 
time were thought to be caused by mixing phenomena during the RGB phase.   
The detection of an abundance spread and element correlations  in main
sequence stars demonstrated that they are the result of pollution from
previous generations of stars  that contaminated the original cluster gas with
CNO (and Na-Mg) cycled material  (Gratton et al. 2001). 

A  vigorous 
discussion is ongoing  about the nature of the stars producing  these
chemical anomalies, the most quoted candidates being either high mass  stars
(Decressin et al. 2007) or intermediate mass AGB's (D' Antona and Caloi 2008).  
Two conclusions  common to all scenarios are that a large amount of He was  
produced and that the polluted stars must be He enriched. 

The discovery of multiple main sequences in several GCs, which
can be  accounted for by a large (sometimes huge) difference in He abundance
among the  different  populations, is  fully consistent with this scenario.   In $\omega$
Cen, where multiple main sequences were firstly  observed,  
the blue main-sequence stars must have a higher He abundance to explain
their higher metallicity (Piotto  et al. 2005).   
A perhaps more remarkable result is that  multiple main sequences are also observed  in NGC 2808 and in other GCs 
without any significant  heavy element abundance spread (Piotto et al. 2007).  
According to stellar evolution prescriptions, large differences in He abundance
are  needed to account for these multiple populations.   In NGC2808, for
example, a detailed photometric and spectroscopic study of RGB bump stars 
requires He abundance differences spanning a range of $\Delta Y$ from 0.11 to 0.19 
among the three sub-populations identified in this cluster (Bragaglia et al. 2010).   
A similar difference, $\Delta Y \sim $0.15, is proposed by D'Antona and Caloi (2008)  
to account for its horizontal branch (HB) morphology. 
From the detailed analysis of several GCs by Gratton et al. (2010), He is most 
likely to be the third parameter, along with metallicity and age, affecting the 
HB morphology.

\section{He abundance variations as diagnostics of multiple populations in GCs}
\label{s:he}

\subsection{The He 10830 line: observations}
\label{s:heobs}

The evidence that He is enhanced in the 'polluted' second-generation stars, is strong
but so far only circumstantial. 

The main obstacle to a direct spectroscopic confirmation is that He is hardly observable 
in cool stars, because no photospheric He lines  are present in their atmospheres. 
Helium lines are observable in hot HB stars of GCs, where however their appearance is 
dominated by other effects, such as diffusion, that prevent  
reliable estimate of abundances (e.g. Mohler et al. 2007). 
The only He lines observable in cool stars are the 10830 {\AA} (and in some 
cases  the fainter 5876 {\AA}, cf. Danks and Lambert 1985) chromospheric lines. 

The 10830 {\AA} line (hereafter He10830) 
has a chromospheric component that forms  in the upper chromosphere 
of a low-gravity cool star farther out than either the H$\alpha$ absorption core 
or the Ca {\sc ii} and Mg {\sc ii} emission cores. 
 The line profile can provide insight into the dynamics of the upper atmosphere, where a wind 
begins to accelerate, hence it has been used with to study mass
loss  and chromospheric activity in several field metal-poor red giants and in a
few  RGB stars in the GC M13  (Dupree et al. 1992; Smith et al. 2004; Dupree et al.
2009).  In these studies  the line was shown to be  detectable in stars 
fainter than  M$_v \sim -1.5$~mag and hotter than $\sim$~4500 K. 
Dupree et al. (2011) observed the He10830 line in twelve 
$\omega$ Cen giant stars belonging to different sub-populations, 
and found variations in the equivalent width as a 
function of the Al and Na abundances, and perhaps also of metallicity. 
They could not, however, quantify these differences in terms of He abundance 
between the different stars. 

Earlier observations of the He10830 line in field stars include the large surveys of
Zirin (1982) and  O'Brien and Lambert (1986).
The latter work is particularly interesting because the authors study in detail 
giants of different spectral types and
the time variability of the line  over 
several timescales.

\subsection{The He10830 line: simulations}
\label{s:hesim}

The He10830 line is actually a triplet spanning the range 9231.8565 - 9230.792 cm$^{-1}$ 
(vacuum wavenumbers), the strongest member of the multiplet occurring at 9230.792 cm$^{-1}$. 
It forms in non-LTE conditions, in the upper chromosphere above 10$^4$ K; 
therefore, it is difficult to use this line to determine the stellar He abundance 
in cool stars,  hence it has never been previously used for this purpose. 
However, we show that {\em relative} He abundances can be estimated with the help of  
adequate chromosphere models. 

We modelled the He10830 line in our target stars based on our previous work 
in NGC 2808 giants (Cacciari et al. 2004), using the Ca {\sc ii} and H$\alpha$ lines 
to model the chromospheres (Mauas et al. 2006). From these studies, it was clear that 
RGB stars in NGC 2808 have structured chromospheres, with plasma as hot as several  
thousands K and, at least for the most luminous objects, sufficiently fast velocity 
fields to produce mass loss along the RGB. 
We therefore expect the He10830 line to be blue-shifted by 
several tens of km~s$^{-1}$ because of the outward velocity field, and to be broadened 
because of the high temperature. 

\begin{figure}
\centering
\includegraphics[width=8cm]{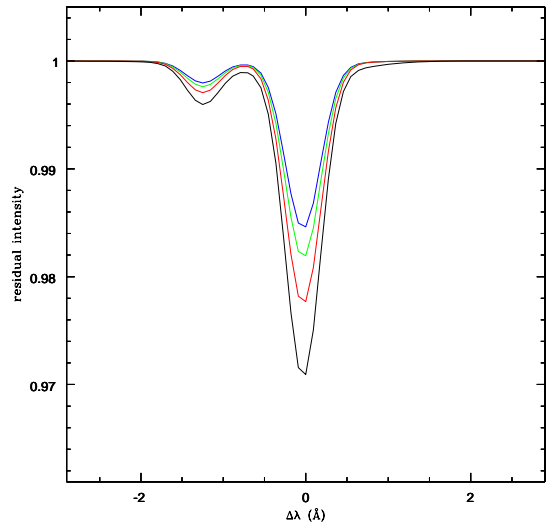}
\caption{Simulations of the He10830 lines for our stars, using the  
chromospheric model of Mauas et al. (2006). 
The line equivalent width changes dramatically with the He abundance: 
the black line corresponds to Y=0.3, and the other lines correspond to a progressive  
decrease in steps of 0.03 in the He abundance. }
\label{f:hel}
\end{figure}

The purpose of these simulations is to investigate how strongly the He10830 line reacts to 
variations in the He abundance and to variations in the chromospheric structure, 
to evaluate the extent to which  the line can be used to distinguish between these two effects. 
Figure \ref{f:hel} shows the He10830 line profiles obtained with our model of NGC 2808 
bright giants (star 48889, Mauas et al. 2006) when varying the content  of He in the atmosphere.
The black line assumes a He abundance Y=0.3, and the other lines are computed by 
decreasing Y progressively in steps of 10$\%$. 

\begin{figure}
\centering
\includegraphics[width=8cm]{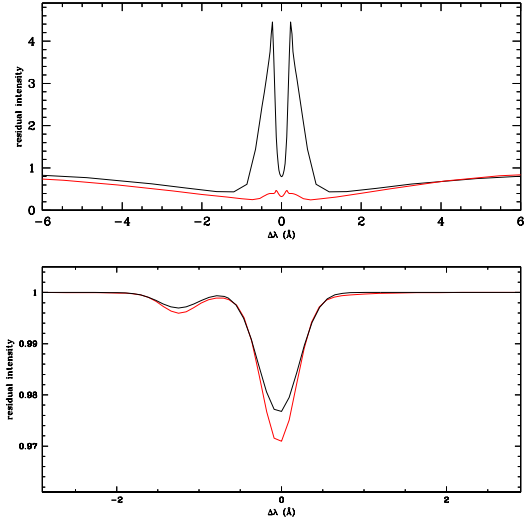}
\caption{Simulations of the He10830 line (lower) and of the Ca {\sc ii} K (upper) line cores 
 for star 48889. The change in the He line strength (corresponding to a lower He abundance 
 by $\Delta$Y = -0.03) is obtained by changing the chromospheric 
 model only.  As a consequence, the emission in the core of the Ca {\sc ii}  line is greatly 
 enhanced, as shown in the upper panel. 
 }
\label{f:helca}
\end{figure}

We also simulate the change in the chromospheric structure  needed to produce  
the same effect on the He line profile as a 10\% lower Y (see Figure \ref{f:helca}). 
The He 10830 line forms in higher layers than the Ca {\sc ii} K line. However, since it 
is dominated by radiation, its source function is decoupled from the 
Planck function in the region where the line is formed, and it depends on the conditions at 
inner layers. As we already pointed out in Mauas et al. (2006), ``the computed profiles 
are unaffected by the structure of the regions where the temperature is higher than 10000 K''. 
Therefore, to mimic the lower He abundance profile the chromosphere has to be heated at layers 
where $T\lesssim 8000$ K. This, however, produces a huge variation on the profiles
of the other chromospheric lines, e.g. the K line that is dominated by
collisions, and its core reflects the temperature
increase much more strongly. This variation would be readily observable, as is
clearly shown in the upper panel of Fig. \ref{f:helca}, where the Ca {\sc ii} K line 
cores obtained with the two chromospheric models are compared.

 This  does not imply that we are able to derive the He abundance from the 
observations of the He 10830 line and its modelling using the Ca II and H$\alpha$ lines. 
However, for two similar metal-poor cool giants, with not only the same main physical parameters, 
but also a similar chromospheric structure, 
as for  our twin target stars, 
it is possible to distinguish between changes in the He lines 
caused by different He abundances or by different chromospheres, because   
chromospheric changes large enough to produce a visible difference in the He 
line profile would also produce huge differences in the other chromospheric 
lines.  We  discuss this point further in Section 4.

\section{Observations and analysis}
\label{s:obsan}

We  selected from the list of Carretta et al. (2004) two RGB stars in  the
cluster NGC~2808, with  very similar stellar parameters but very different   Na
and O abundances (see Table \ref{t:targets}).   The Na-rich O-poor star has
been polluted by processed gas from a previous generation of stars and is expected 
to be representative of the He-enhanced population.  
The other  star, with standard O and Na values, is taken to be representative of 
the unpolluted  He-normal (primordial) population.

The observations were carried out in service mode at the ESO VLT.
The UVES spectrograph (Dekker et al. 2000) was used to observe the stars in 
the  Ca {\sc ii} H and K lines and in H$\alpha$ at R $\sim$ 40000. 
The IR  high resolution spectrograph CRIRES (K\"aufl et al. 2004) was used for the He lines,  with a similar resolution.

Complete information about the target stars and the observations is contained 
in  Table 1.  
The temperature  values are those based on B-V because no V-K colour is available for 
star 46422. 
The two stars have the same parameter values in the spectroscopic analysis of Carretta
et al. (2004), but the spectroscopic temperatures are 100 K higher than the values based 
on the B-V colours.

Our stars
are slightly brighter than M$_V$ = --1.5 mag and cooler than 4600 K, which represent
the He10830 line detection thresholds found by Smith et al. (2004) and
Dupree et al. (2009), so it was not quite clear whether the He line
would be detectable in the normal abundance star.  On the other hand,
these stars are bright enough to permit a spectrum  of the appropriate signal-to-noise
ratio (S/N) to be obtained in
a reasonably short integration time.

The spectra were processed starting from the output of the ESO pipelines.
The UVES spectra taken in two separate runs (early November and December) are very 
similar for each star, therefore the average spectrum was used. 
We  also observed  a pair of fainter and hotter stars of NGC 2808, but the S/N of the
extracted CRIRES spectra was not high enough to permit  a reliable analysis;  improvements will require the 
modifications of the pipeline extraction software. 

The expected 
depth of the He line from our models is fairly small, 
therefore,  even if the S/N in the continuum of the stars is around 100 
for a 0.044 cm$^{-1}$ pixel,
for the two brightest stars a special care was taken and 
the output spectra of the CRIRES pipeline (Jung \& Bristow 2008 ) were further treated; 
all spectra were deglitched and normalized to unity for the continuum using a low order polynomial.

The region around the He10830 lines contains two telluric lines, a strong one at 9229.284 cm$^{-1}$, 
and a weaker one (about 1/3 of the equivalent width) at 9227.689 cm $^{-1}$ (Breckinridge and Hall 1973). 
The presence of these telluric  lines can endanger a correct detection or measurement  of the He10830 lines.
We  therefore observed a rapidly rotating hot star for telluric
correction (HD~79041, B6V, $v sin(i) = 248 km/s$), at an airmass similar to the one of the cluster targets (1.5 for HD~79041, 
1.7 for 48889 and 1.53 for 46422). 
The spectrum of the standard star was first normalized and
scaled by the ratio of the airmasses, to properly reproduce the depth of the telluric lines.
The program star spectra were divided by the scaled spectra  of the standard star. 
Figures 3 and 4 show in black the spectra of the program stars and in red the spectra of the standard star  before the division. 
In these figures, wavelength is expressed in cm$^{-1}$, no velocity adjustment has been applied and blue is to the right. 
The reader can note in Figures 3 and 4 that the telluric lines 
are redshifted by $\sim$ 0.5 cm~s$^{-1}$ with respect to their  nominal wavenumber
because of the imperfect wavelength calibration. 

When taking into account the 
radial velocity of the stars (cfr. Table 1), the He10830  lines are expected to be blueward of  
9227.25 cm$^{-1}$ for star 48889 and of 9227.92 cm$^{-1}$ for star 46422.

\begin{figure*}
\centering
\includegraphics[width=18cm, height=7cm]{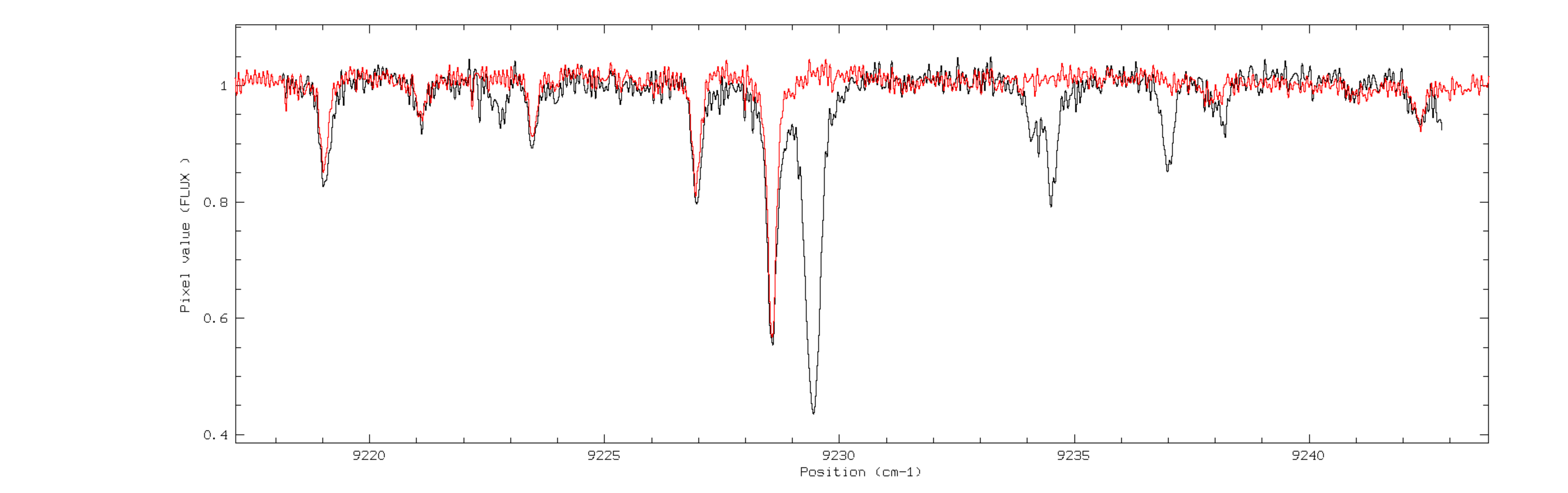}
\caption{Spectra of the star 48889 (black) and of the telluric standard star (red) before telluric correction. The abscissa is in cm$^{-1}$, 
red is left. The wavelength calibration is from the CRIRES pipeline, without further adjustment, and shows a shift of about 0.5 cm$^{-1}$ 
towards the red  with respect to the  nominal wavenumber of the telluric lines. The wavelength calibration and the correction for stellar 
radial velocity were performed in the last step of the data reduction, after telluric correction.}
\label{f:48889nocorr}
\end{figure*}

\begin{figure*}
\centering
\includegraphics[width=18cm, height=7cm]{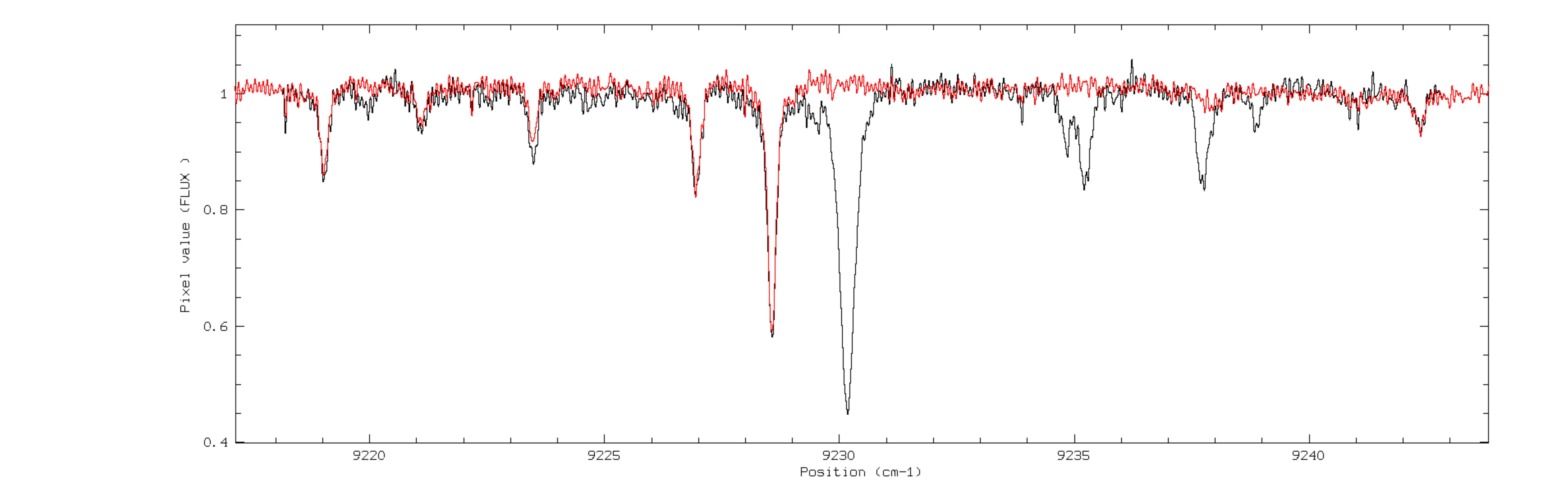}
\caption{Spectra of   star 46422 (black), and of the telluric standard star (red) before telluric correction.The abscissa is in cm$^{-1}$, 
red is left. The wavelength calibration is from the CRIRES pipeline, without further adjustment, and shows a shift of about 0.5 cm$^{-1}$ 
towards the red  with respect to the  nominal wavenumber of the telluric lines. The wavelength calibration and the correction for stellar 
radial velocity has been performed in the last step of the data reduction, after telluric correction }
\label{f:46422nocorr}
\end{figure*}

The last step consisted of correcting  the program star
spectra for radial velocity  
by matching the McMath Solar FTS spectra
using the Solar FTS atlases from Wallace et al. (1993) centred on the HeI
line. A small adjustment to the wavelength scale was also required to best match the 
solar atlas. The finally reduced spectra at rest-frame wavenumber  are shown in Figure 5. 
The spectra of the two NGC~2808 stars, show the same  features, 
with the exception of two lines: at $\sim$ 9227 cm$^{-1}$, where a Na
doublet is present in star 48889 and is much smaller in star 46422, thus confirming the
much higher Na abundance of star 48889; and at $\sim$ 9231.32 cm$^{-1}$, where
a rather broad line is clearly present in the spectrum of star 48889 but is
missing in the spectrum of star 46422.

This feature has a different shape from the other lines:  it is at least  17 km~s$^{-1}$  broad, and is 
blue-shifted by $\sim$17 km~s$^{-1}$  with respect to the rest wavelength 
of the strongest He10830 component (9230.792 cm$^{-1}$).  
This is approximately what is expected,   
in a few field stars the He lines  have blue shifts larger than 60 Km~s$^{-1}$ (Dupree et al. 2009), 
but most field giants have spectra strikingly similar to our spectrum of star 48889.
Several spectra presented in O'Brien and Lambert (1986) also have He lines
that are similar in depth and width to the one observed in star 48889.    
The wavenumber of the nearby Si (9233.58) and Na (9226.94) lines are in excellent
 agreement with their vacuum measurements.  

Since the expected (and the observed)  He line is quite small (a few $\%$ of the continuum only), it is necessary to 
further develop  some considerations about the significance of its detection. 
The S/N in the continuum for these spectra is  S/N$\sim$100 for 0.044 cm~s$^{-1}$ pixel for both stars. 
For such a S/N, the error associated with the equivalent width  of the He line is about 4 mcm$^{-1}$, or seven times smaller than  the measured 
equivalent width (Cayrel 1988; see also discussion in Bonifacio et al. 2002).
Apart from the contamination by the 
telluric line, there is, therefore  no reason to question the reality and the detection of the He line. This is also confirmed by  
equivalent widths as small as a few mcm$^{-1}$ being clearly observed and measured in the spectra (cf. Figures 3-5 and Table 2).
The only doubt that can therefore be reasonably cast concerns the blending of the telluric 9227.799 line with  the He feature. 
However, as is clear from Figure 3, the telluric and the He line are partially detached; 
in addition, we  performed several tests 
 changing the intensity of the telluric line, which did not  substantially modify  the equivalent width of the He line. 
We also  note that the telluric lines have a width of 0.22 cm~s$^{-1}$, while the He line is at least 0.52 cm~s$^{-1}$ wide. 

To avoid biases in our analysis, we asked  a colleague to 
measure the equivalent width of the lines in the spectra 
and  use an automatic spectral 
analysis program, DAOSPEC (Stetson and Pancino 2008). 
Among other features, DAOSPEC defines its own continuum.   
A list of line identifications, measured wavenumbers, and equivalent widths for this region 
is given in Table 2. The equivalent widths are an average of the manual and  the DAOSPEC measurements. 
The largest difference  between these measurements was about 6 mcm$^{-1}$, but for the strong Si line, for which the difference is up to 20 mcm$^{-1}$. 
DAOSPEC also estimates as well the errors in the  equivalent widths, which are typically   1.5-2 
mcm$^{-1}$, perfectly compatible with the measured differences. 

There is good agreement between the equivalent widths of the lines of the two stars, 
but most lines  appear slightly stronger in the spectrum of star 48889. This likely indicates that 48889 is slightly cooler than 46422. 
The line at 9231.3 is absent in the spectrum of star 46422. We also  note that 
all the other lines have line  widths not larger than 12 Km~s$^{-1}$ (0.4 cm$^{-1}$), while the 9231.3 feature is substantially wider.
Close to the 9231.3 line towards the blue there is a line at $\sim$9232.88 cm$^{-1}$ that is stronger in star 48889. 
This line could  in principle be identified with a Si line at 9233.10 cm$^{-1}$, but the frequency of the transition 
agrees  very well with a Ti line at 9232.83 cm$^{-1}$. 
This line is sometimes seen in absorption in the spectra of cool M stars (O'Brien and Lambert 1986). 
O'Brien and Lambert (1986) also mention  an unidentified line at 
10831.3 {\AA} (or 9229.975 cm$^{-1}$)
in the spectra of stars later than spectral class M, with an equivalent width of 16 m{\AA} in  $\alpha$Her.
This line is visible in star 46422 with an equivalent width 
of 9 mcm$^{-1}$, and we can set an upper limit of 6 mcm$^{-1}$ for star 48889; our targets are hotter 
and fainter than $\alpha$Her and in addition  have a rather low metallicity, which would naturally decrease the strength of most lines.

Finally, we  consider the possibility that the 9131.3 feature is the blend of two so far unidentified lines. In principle, it can be 
reproduced with two lines, one at $\sim$9131.15 and one at $\sim$9131.6 cm$^{-1}$, 
each with an equivalent width of about 15 mcm$^{-1}$. 
We are not aware of known or reported lines at these  wavenumbers; 
possible nearby candidates are two Si lines (9131.81, 9131.32). 
However, we note that the difference between the two stars for the two lines would be very large, whereas a few Si lines we checked in the UVES
spectra (6125.021, 6142.483 {\AA}) agree to within 2 m{\AA} in their equivalent widths. In 
addition, if the observed feature were the product of the blend of the two Si lines, one may also 
expect companion lines at 9130.42 and  at 9233.10 cm$^{-1}$, 
which are not observed ( as discussed above; note that 
the bluer would be  partially blended with the strong 9233.56 line). 

The combination  of the large equivalent width of the line, 
 the missing companion Si lines,  that no lines have been reported at this wavenumber in the literature 
(even for more metal rich and/or cooler stars) and that the 
observed feature  qualitatively agrees with what is expected a priori, 
let us conclude that the 9131.3 line is due to He, although we cannot exclude 
a priori that it is  a blend of two unknown lines.  


\section{He abundance difference}\label{s:hediff}

We  estimated the  difference in He abundance using our chromospheric models. 
The Ca {\sc ii} K line core profiles in Figure 6 show that the star 48889 was 
slightly more active than star 46422, 
but the difference is not so dramatic as to require the use of a completely different 
model. This difference is compatible with a low level of chromospheric variability.

The basic  models are those for star 48889 developed in Mauas et al. (2006), which have
only been modified in the  upper chromosphere and made slightly hotter to take into account 
the higher chromospheric activity level of this star in 2010  with respect to 2004. 
However, this update does not involve a substantial change from the previous model. No
velocity fields are considered for these simulations.  
Maintaining the same T versus (vs.) log(m) distribution, we self-consistently
recomputed the non-LTE radiative transfer and the statistical and
hydrostatic equilibrium equations, assuming different values of the He
abundance. Using the program Pandora (Avrett \& Loeser 1984), in
each case we computed non-LTE populations for 10 levels of H, 9 of C {\sc
i}(see Mauas, Avrett \& Loeser 1990), 15 of Fe
{\sc i}, 8 of Si {\sc i} (see Cincunegui \& Mauas 2001), 8 of Ca {\sc i} and Na {\sc i},  
6 of Al {\sc i} (see Mauas,
Fern\'andez Borda \& Luoni 2002), and 7 of Mg {\sc i} (see Mauas,
Avrett \& Loeser 1988). In addition, we computed 6 levels of He {\sc
ii} and Mg {\sc ii}, and 5 of Ca {\sc ii}. For every species under
consideration, we included all the bound-free transitions and
the most important bound-bound transitions.The Ly-$\alpha$,
Ca {\sc ii} H and K, and Mg {\sc ii} $h$ and $k$ lines were all
computed with a full partial-redistribution treatment (for a
discussion, see Falchi \& Mauas 1998).
For He {\sc i}, we used the 29-level atomic model used by Mauas et al. (2005) and Andretta et
al. (2008).

It is well known that a strong coronal radiation can ionize He, and
  enhance the strength of the line via recombination to its higher
  levels (for a discussion of this mechanism in the Sun, see Mauas et
  al. 2005). However, these stars are well beyond the so-called
  coronal 'dividing line' (Linsky and Haisch 1979, Maggio et
  al. 1990), and are therefore by far too cool and too luminous to
  possess a corona. The only coronal activity detected in
  metal poor stars is in short-period binary systems (Pasquini et
  al. 1991), which is not the case for our sample stars.
  Furthermore, O'Brien and Lambert
  (1986) found that this effect is not important for giant stars.

The resulting spectra were broadened by 
a Gaussian of 17 Km~s$^{-1}$  FWHM.

Figure 7 shows the results of the modeling: star 46422 spectrum is compatible with a non 
detection. 
 The figure shows the model with Y$\sim$0.22, that has an equivalent width close to 
the upper limit in Table 2. We consider this value a valid upper limit and the model a 
good representation of the observed spectrum. An extra absorption in this area 
seems to be present, so at face value a model with no He line would not be fully compatible with the spectrum.
However, given the uncertainties in the He line width and the critical placement of the continuum, 
 a claim of He detection for this star would be an overinterpretation of the data.

For star 48889, an abundance of Y$\ge$0.39 is needed to reproduce the observed spectrum. 
We have tried different values of Y, and the model line profile would still be compatible 
with the observed one for a He abundance as high as Y$\sim$0.5 in this star. 
When considering all the uncertainties involved, as discussed in the previous 
section and the uncertainty in the placement of the continuum,
we infer  that our Y=0.39 model is a lower limit for this star. 
Since we aim to derive relative He abundances rather than absolute values that 
are affected by much larger uncertainties, 
our models show that a difference of at least $\Delta$Y=0.17  
is needed to reproduce the observed spectra. 

Such a large difference in He abundance would increase the opacity, therefore modifying   
the spectra of  these He-rich stars. This problem was studied  by 
Bohm Vitense (1979), who pointed out that the effect is an increase in the gravity 
of the stars,  as quantitatively addressed by 
Pancino et al. (2011). A helium abundance increase by $\Delta$Y=0.10 (from 0.25 to 0.35) 
in subgiant stars leads to a higher metallicity by $\Delta$[Fe/H]=0.08 dex, 
based on a simplified treatment of the atmosphere where the increased helium abundance 
is approximated by an artificial increase in the gravity. 
A more refined atmospheric modelling finds a smaller metallicity increase, i.e. $\Delta$[Fe/H] 
$\simeq$ 0.02-0.03 dex (Pancino, private communication).
If we apply the Gray (2008) approximation to our star, a helium increase of $\Delta$Y=0.17 would correspond to a $\Delta$\logg=0.06, 
that in turn leads to increase the metallicity of about 0.01 and 0.03 dex for FeI and FeII, respectively (Carretta, private communication).   
This difference is much smaller than the typical error in the metallicity determination of 0.10-0.15 dex estimated by Carretta et al.
(2004), who derive essentially the same values of [Fe/H] within the errors for the two stars (see Table 1).   

As we pointed out in Section 2.2, it is not possible to modify
  the He profiles by changing only the upper layers of the atmosphere.  
Modifying the atmosphere above around 9000 K does not change
  the He profiles, because the source function of the lines is completely
  determined by the conditions at lower heights. This  was
  mentioned in Mauas et al. (2006), where we pointed out that,
  although the center of H$\alpha$ is formed at around 12000 K (see
  Fig. 2 of Mauas et al. 2006), the computed profiles are unaffected
  by the structure of the regions where the temperature is higher than
  10$^4$ K. The profile of the He 10830, which as in the case of H$\alpha$ is dominated by
  radiation, is also unaffected by changes high in the atmosphere. 
  On the other hand, the chromospheres of these stars are, of course, far from
  being semi-static and homogeneous, as our models assume. We do not
  have a realistic way to evaluate the inhomogeneities of the
  chromosphere of these stars in any meaningful way. However, to
  the best of our present knowledge, we believe that a He
  abundance difference is the simplest and most plausible explanation
  of the observational evidence.

\begin{table*}[t]
\caption{Target stars. Astrophysical parameters and log of observations.}
\label{t:targets}
\begin{center}
\begin{tabular}[t]{lccccccccc}
\hline\noalign{\smallskip}
Star name   &  \Teff & \logg   & M$_v$ & [Fe/H] & [Na/Fe] & [O/FE] & RV         & Obs. UVES            & Obs. CRIRES  \\
            &        &         &       &       &          &        & km~s$^{-1}$ & int. time = 1200 sec & int. time = 1800 sec \\
\noalign{\smallskip}
\hline\noalign{\smallskip}
48889     & 3843  & 0.52  & -2.25 & -1.22 $\pm$0.12& +0.64   & -0.07       & 115.1      &  06/11/09 \& 26/12/09 & 26/11/09  \\
46422     & 3843  & 0.42  & -2.21 & -1.08$\pm0.09$ & +0.12   & +0.23       & 93.5       &  06/11/09 \& 02/12/09 & 26/11/09  \\
\noalign{\smallskip}
\hline
\end{tabular}
\end{center}
\tablefoot{Stellar identification and astrophysical parameter values are from Cacciari et al. (2004) and Carretta 
et al. (2004). The [Fe/H] quoted are the values obtained from  FeII.
}
\end{table*}

\begin{table*}[t]
\caption{Spectral lines in the spectral region around the He10830 line }
\label{t:lines}
\begin{center}
\begin{tabular}[t]{ccccccc}
\hline\noalign{\smallskip}
Wavenumber (Meas.)   &  Wavenumber (Exp.) & Element   & Width  (48889) & Width (46422)              & Eq. Width (48889) & Eq. Width (46422)    \\
       cm$^{-1}$     &    cm$^{-1}$        &           &  $\Delta$cm$^{-1}$   &    $\Delta$cm$^{-1}$   &  mcm$^{-1}$            &    mcm$^{-1}$        \\
\noalign{\smallskip}
\hline\noalign{\smallskip}
9219.27 & 9219.29 & Si I & 0.33 & 0.35    & 89     & 85   \\
9223.34 & 9223.44 & Ca I  & 0.37 & 0.42  &  23   &  17 \\
9225.99 &  ?      & ?     & 0.2  & 0.40  &  8 &   9   \\
9226.94 & 9226.90 & Na I & 0.35  & 0.41 &   28   &  12 \\
9229.97 & ?       &  ?   &  /    & 0.24 &  $<$ 6 & 9  \\
9231.30 & 9230.792 & He I & $>0.52$  &  /    &  30  & $<$11  \\
9232.88 & 9232.83 & Ti I & 0.38  &   0.43  & 36  & 27  \\
9233.58 & 9233.56  & Si I  & 0.40  & 0.41  & 252    & 240 \\
9238.25: & 9238.17  & Cr I  &  0.31  & 0.29  & 36:  & 30  \\
9238.60: &  ?      & ?  & 0.30  & 0.30  & 62  & 54  \\
\noalign{\smallskip}
\hline
\end{tabular}
\end{center}
\tablefoot{ Table with the measured absorption  lines in the spectra of the two NGC2808 stars in the region shown in Figure 3. The 
measured wavenumber (column 1) and the expected wavenumber (column 2), as well as the line width for the two stars (columns 4 and 5)
are given in cm$^{-1}$. The equivalent widths (columns 6 and 7) are in mcm$^{-1}$. For all lines, a gaussian approximation was used. }
\end{table*}

\begin{figure}
\centering
\includegraphics[width=10cm, height=9cm ]{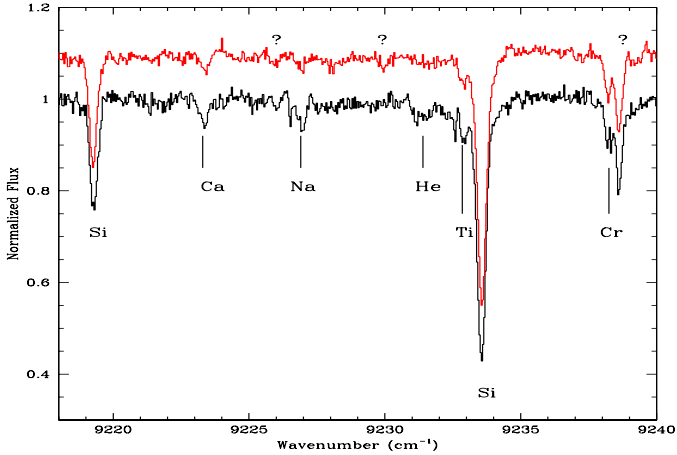}
\caption{Spectra of the targets:  star 46422 (top - red), star 48889 (bottom - black). 
All the lines listed in Table 2 are indicated. The question mark indicates 
lines measured but with no element identification.  }
\label{f:3spectra}
\end{figure}

\begin{figure}
\centering
\includegraphics[width=7.0cm, angle=270]{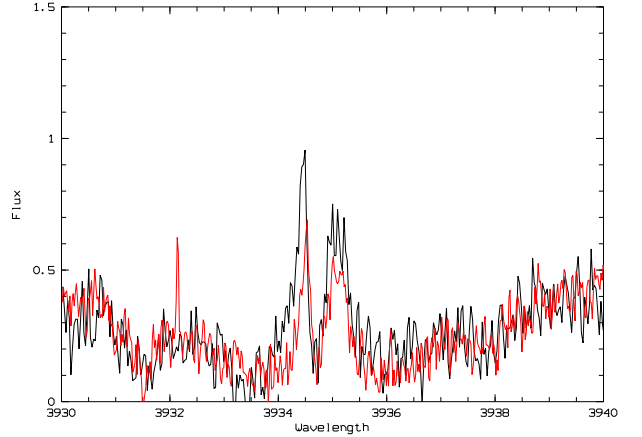}
\caption{Ca II K line core spectra of the targets; star 48889 (black) is slightly more 
active than star 46422 (red), as the core emission is slightly higher.
}
\label{f4:caii}
\end{figure}

\begin{figure}
\centering
\includegraphics[width=10cm, height=9cm,  angle=0]{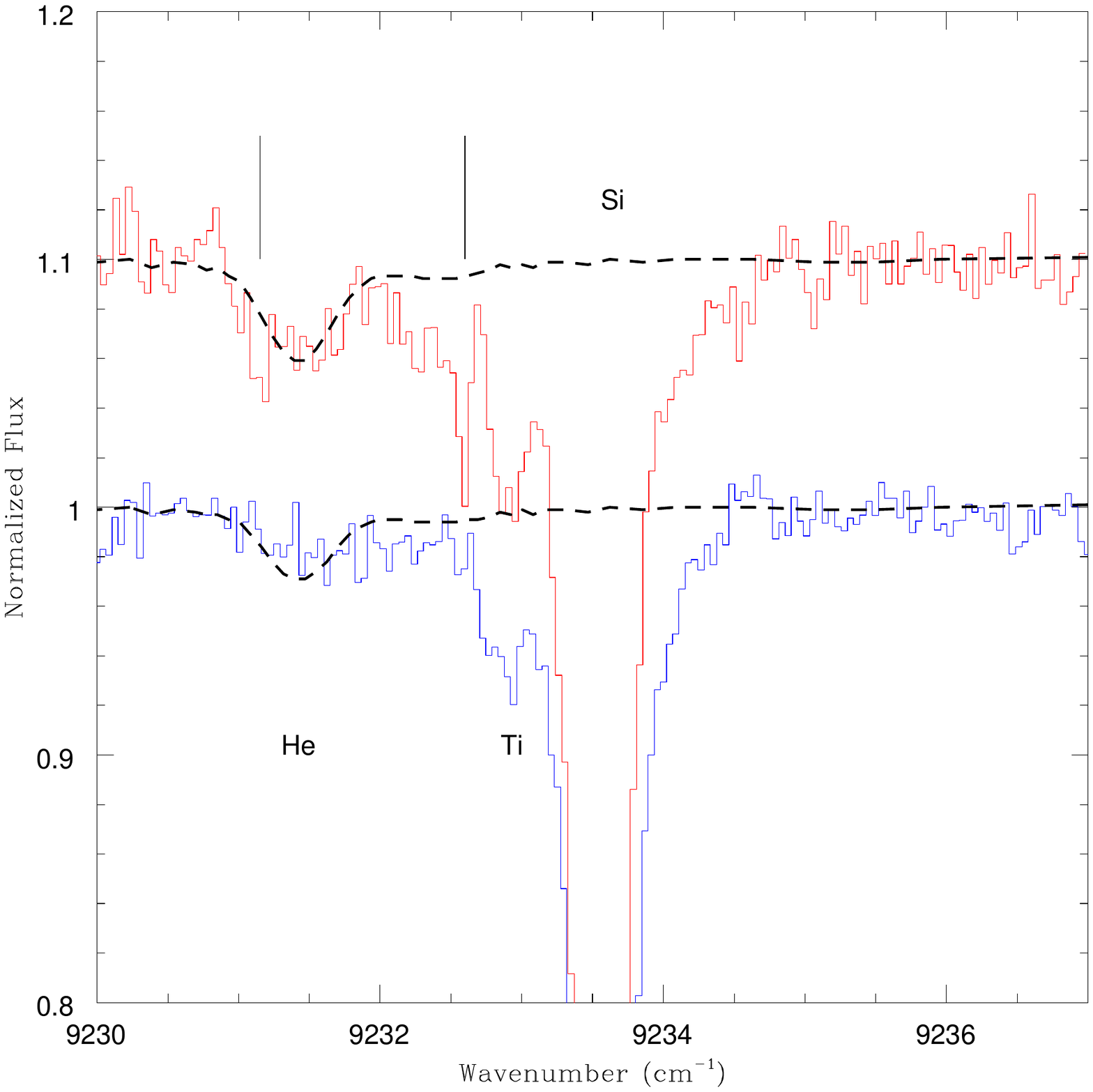}
\caption{Observed spectra vs. He models. 
The He lines in the models have been shifted by 17 km/s. The model superimposed on the data for  star 
46422 (lower spectrum) has Y=022. 
We find that this model is the most He-rich still compatible with the observed 
spectrum (see text).  
For star 48889 (upper spectrum), the model  has Y=0.39. 
This is the minimum He abundance compatible with 
the observed spectrum, but a higher value of Y cannot be excluded.
Observed spectra are continuous lines, where the models are dashed (black) lines. Other spectral 
lines are marked. 
The position of the telluric lines in the spectrum of star 48889 is marked 
as vertical bars above the spectra.  }
   
\label{f4:caii}
\end{figure}
 
\section{Summary and conclusions}
\label{s:end} 

We have studied two RGB stars in the 
GC NGC 2808, that have
 very similar astrophysical parameters but very different Na and O abundances. 
These stars are assumed to represent the initial sub-population with primordial Na, O, and He 
abundances, and a subsequent subpopulation formed out of processed material from the previous 
population, which is  hence Na and He enriched and O depleted.
Whereas Na and O abundances are available for these and other stars in NGC 2808, the He abundance 
has never been determined before. 
Since helium differences are supposed to be responsible for peculiar observed features of the 
CMD such as the multiple main sequences and the HB morphology, the empirical detection of 
a He difference would provide the essential and final confirmation of this scenario. 

Helium is barely detectable in cool stars, the only observable line being the He10830  
chromospheric line. This line forms in non-LTE conditions and should not be used  to 
derive absolute He abundances, although with the help of a suitable chromospheric model, 
reliable quantitative information about relative abundances can be obtained under appropriate 
assumptions.  
Our study relies on detailed  model chromospheres based on high-resolution 
VLT-UVES spectra of the Ca {\sc ii} K and H$\alpha$ lines for the two program stars. 
These models produce the corresponding He10830  line profiles for comparison with 
the observed ones, obtained from VLT-CRIRES high-resolution spectra. 

The two program stars have distinctly different 
features in the spectral region blueward of the He line rest frequency. In particular an absorption line,
broader than the other recognized lines, in the spectrum of star 48889, which is a sodium-rich, oxygen-poor star) 
is absent in the spectrum of star 46422 (which has normal O and Na abundances). 
We interpret this feature as a blue-shifted He line and use our models to derive the He abundance differences.
The He abundance difference estimated from 
the models is no less than $\Delta$Y = 0.17, in excellent agreement with theoretical 
predictions based on stellar evolution considerations to account for the observed 
peculiarities of the CMD. 
Our results and conclusions are based on the assumption that the two observed 
stars are so similar, including  in terms of their chromospheres, that the same model can 
be applied to both of them. This assumption relies on there being a  strong similarity 
between their photometric and spectroscopic characteristics.  We
  believe that an abundance difference between both stars is the most
  natural explanation of the difference in the observed profiles
 
As a final remark, we also assume that there is no variability between the 
sets of the infrared and the bracketing optical measurements, 
hence the chromospheric characteristics that can be modelled by 
the Ca {\sc ii} and H$\alpha$ lines apply also to the He10830 line. 
However, variability on different timescales in the He10830 line was observed
in several field giants, e.g. by Lambert and O'Brien (1986), as pointed out 
by the referee.  Although it cannot  in principle be excluded that a 
significant chromospheric variation may have occurred in star 48889 at the 
time of the He10830 observation, which was not monitored by simultaneous optical 
observations, this variation should not have   affected the
Ca {\sc ii} and H$\alpha$ spectra that bracket the He10830 
spectrum by just -20 and +30 days. This seems rather unlikely to us, and we 
feel  confident that the enhanced He10830 line in star 48889 is 
indicative of higher helium abundance rather than of variability. 
However, future observations should take this possibility into account.  
The detection of He in GC giant stars still has a very short history, and clearly 
it will be important to measure the extent to which and on which timescales 
He10830 variability occurs in these objects. 

\begin{acknowledgements}
The authors gratefully acknowledge the use of 
NSO/Kitt Peak FTS data, produced by NSF/NOAO. 
We thank E. Pancino and E. Carretta for useful comments on metallicity determination in He-enhanced atmospheres.
The comments and suggestions of an anonymous referee contributed to improve the quality of the paper.  
Rodolfo Smiljanic kindly performed the independent measurements of the equivalent widths and suggested 
 that the 9231.3 line might be a blend.
\end{acknowledgements}


{}

\end{document}